\documentclass[%
 reprint,
superscriptaddress,
 amsmath,amssymb,
 aps,
]{revtex4-1}

\usepackage{graphicx}
\usepackage{dcolumn}
\usepackage{bm}

\begin{document}

\preprint{APS/123-QED}

\title{Broadband enhancement of the magneto-optical activity of hybrid Au loaded Bi:YIG}

\author{Spiridon D. Pappas}
 \email{pappas@rhrk.uni-kl.de}
\author{Philipp Lang}
\author{Tobias Eul}
\author{Michael Hartelt}
\affiliation{Fachbereich Physik and Forschungszentrum OPTIMAS, Technische Universit\"at Kaiserslautern, 67663 Kaiserslautern, Germany}
\author{Antonio Garc\'{\i}a-Mart\'{\i}n}
\affiliation{Instituto de Micro y Nanotecnolog\'{\i}a IMN-CNM, CSIC, CEI UAM+CSIC, Isaac Newton 8, E-28760 Tres Cantos, Madrid, Spain}
\author{Burkard Hillebrands}
\author{Martin Aeschlimann}
\author{Evangelos Th. Papaioannou}
\affiliation{Fachbereich Physik and Forschungszentrum OPTIMAS, Technische Universit\"at Kaiserslautern, 67663 Kaiserslautern, Germany}

\date{\today}

\begin{abstract}
We unravel the underlying near-field mechanism of the enhancement of the magneto-optical activity of bismuth-substituted yttrium iron garnet films (Bi:YIG) loaded with gold nanoparticles. The experimental results show that the embedded gold nanoparticles lead to a broadband enhancement of the magneto-optical activity with respect to the activity of the bare Bi:YIG films. Full vectorial near- and far-field simulations demonstrate that this broadband enhancement is the result of a magneto-optically enabled cross-talking of orthogonal localized plasmon resonances. Our results pave the way to the on-demand design of the magneto-optical properties of hybrid magneto-plasmonic circuitry.





\end{abstract}

\pacs{Valid PACS appear here}
\maketitle


The field of magneto-plasmonics has attracted a lot of scientific research, both for its importance in potential technological applications, and for its fundamental scientific interest \cite{Magnetic_plasmonic_functionalities,Ignateva,bio-sensing15,Antonio2016}. Towards these directions, various ideas like using external magnetic fields to control the dispersion relation of plasmonic resonances (active magneto-plasmonics) \cite{active_magnetoplasmonics}, or using plasmonic resonances for spin current generation in adjacent magnetic insulators \cite{Uchida2015,Tadaaki17}, and the implementation of a magneto-plasmonic interferometer \cite{Elezzabi16}, have been realized and explored.

Magneto-optical studies on magneto-plasmonic nanostructures, composed of magnetic and/or plasmonic materials, have revealed new exciting effects: Nanopatterned hybrid heterostructures~\cite{PhysRevLett.104.147401,papa2010,Maccaferri2015,Papaioannou15, PhysRevB.88.075436,doi:10.1021/acsphotonics.6b00670,Caballero:15,martin-becerra:183114}, pure ferromagnetic films \cite{Hui15,Kataja15, melander:063107,PhysRevB.93.214411,doi:10.1021/nl2028443,Papaioannou11,LUONG201879,PhysRevLett.111.167401}, and noble metal/magnetic dielectric systems \cite{Belotelov2011,PhysRevLett.96.167402} exhibit plasmon-induced enhancement of their magneto-optical activity. Recently, the correlation of near- and far-field effects of a patterned magneto-plasmonic array has been achieved with the aid of a Photoemission Electron Microscopy (PEEM)~\cite{doi:10.1021/acs.nanolett.5b05279}, paving the way for tailoring the magneto-optical response of these systems: The spatial distribution of the polarization- and energy-dependent electric near-field of the propagating plasmon polaritons has been connected to the size of the enhancement of the magneto-optical Kerr effect. As a further matter, the nature of the plasmonic resonances (localized or propagating surface plasmons~\cite{PhysRevB.74.245415,Papaioannou:17}) can alter differently the response of the magneto-optically active material in a magneto-plasmonic structure. From the wider family of the hybrid magneto-plasmonic structures~\cite{doi:10.1021/acsphotonics.6b00670,Caballero:15,martin-becerra:183114,Borovkova:16,Belotelov2011, Maksymov:14}, the system composed of ferrimagnetic dielectric layers of bismuth-substituted yttrium iron garnet (Bi:YIG) with embedded Au nanoparticles (AuNPs) that support localized plasmon resonances (LPRs), has attracted little attention for its importance~\cite{PhysRevLett.96.167402, Almpanis:16}. Furthermore, the underlying near-field mechanism is not clarified yet, while there is lack of studies on the longitudinal MOKE (L-MOKE) configuration induced by LSPs.

In this letter, we explore the influence of embedded self-assembled AuNPs that support LPRs, on the magneto-optical activity of a host Bi:YIG layer. The experimental results show a broadband enhancement of the magneto-optical activity of the surrounding Bi:YIG in the spectral region where the plasmonic resonances occur. The analysis of the simulated near- and far-field behavior reveal the existence of two orthogonal localized plasmon resonances, attributed to each of the optical semi-axis of the nanoparticles. We show that these plasmon resonances are coupled through the magneto-optical properties of the host material, creating a broadband magneto-optical enhancement.

\begin{figure*}
  \includegraphics[width=0.8\textwidth]{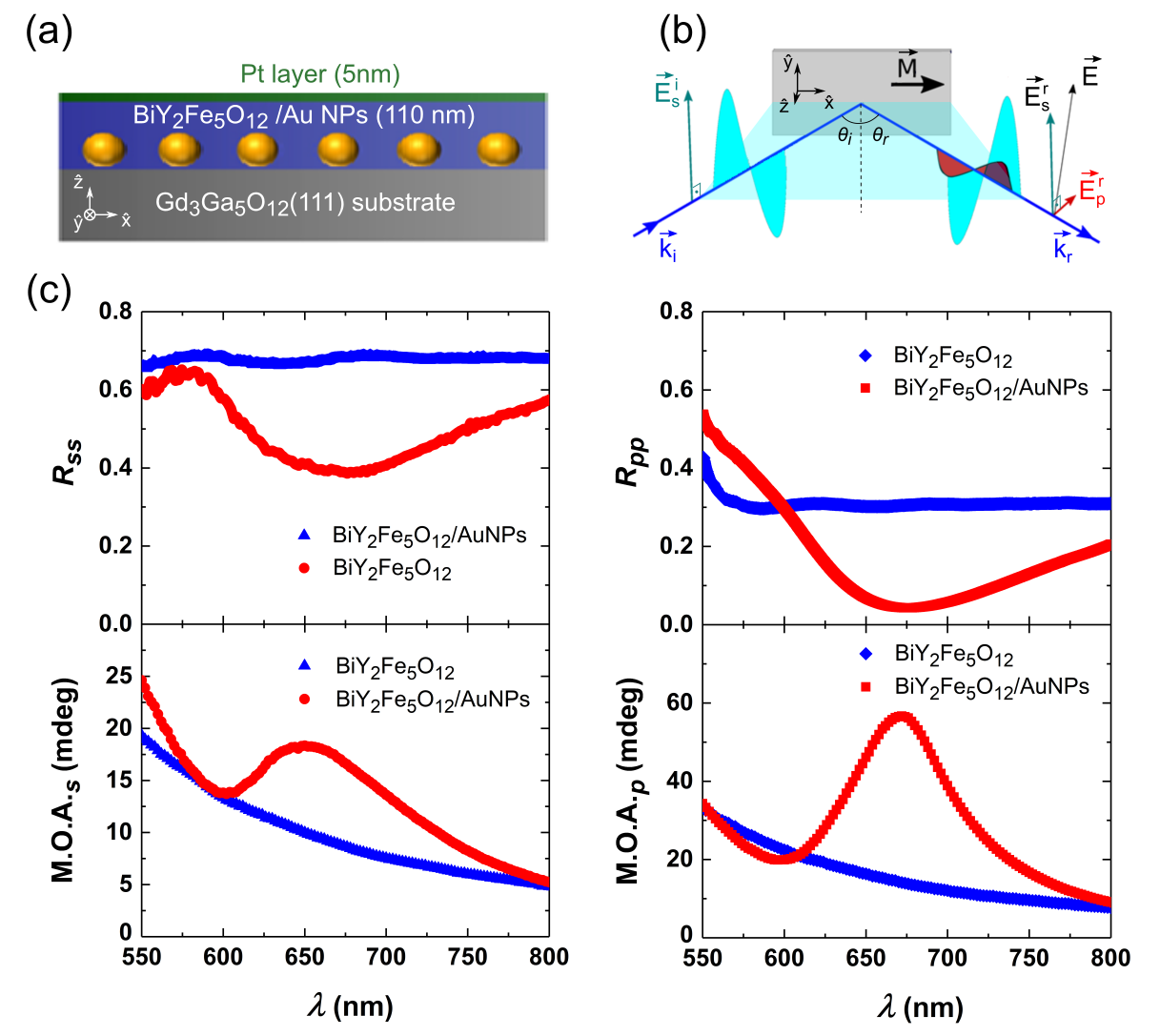}
  \caption{(a) Schematic illustration of the sample used in the present study. (b) Scheme of the longitudinal MOKE configuration, used to record the Kerr angle of the BiY$_2$Fe$_{5}$O$_{12}$/(AuNPs) samples. The Kerr angle response of the samples is recorded for a range of different wavelengths of incident light. (c) (Top graphs) Experimental $R_{ss}$ and $R_{pp}$ reflectance plots as a function of wavelength for the BiY$_2$Fe$_{5}$O$_{12}$/(AuNPs) samples. An aluminium mirror was used as a reference. (Bottom graphs) Dependence of the magneto-optical activity ($\textrm{M.O.A.}_{s(p)}$) as a function of wavelength, recorded for the BiY$_2$Fe$_{5}$O$_{12}$/(AuNPs) samples for s- and p-polarized incident light. Both the reflectance as well as the M.O.A. spectra have been recorded for angle of incidence $\theta_{i} = 55$~degs.}
  \label{Figure1}
\end{figure*}

In Fig.~\ref{Figure1} (a), the structure of the investigated hybrid sample (Bi:YIG/AuNPs) is presented with the aid of a sketch. Scanning electron microscopy (SEM) images of the sample (see Ref. [11]) reveal that the AuNPs are randomly distributed close to the interface between Bi:YIG and gadolinium gallium garnet (GGG), and that they are embedded in the Bi:YIG film. Furthermore, they have the shape of oblate spheroids, with an in-plane diameter ranging from 30 to 90 nm and a height ranging from 20 to 50 nm. The optical properties of this sort of system is mainly determined by localized plasmon resonances (LPRs), rather than by geometrical-lattice resonances. A second sample, having exactly the same structure, but containing no AuNPs (BiY$_2$Fe$_{5}$O$_{12}$), was used as a reference. The reader is referred to the Supplemental Material for further details on the fabrication of the samples.

The magneto-optical response of the samples was experimentally probed in the L-MOKE configuration. The latter is schematically depicted in Fig.~\ref{Figure1}(b). In the L-MOKE geometry, the magnetization ($\overrightarrow{M}$) of the sample lies in the sample plane, as well as in the plane of incidence of the incident light. The incident light can either be s- or p- polarized. In Fig.~\ref{Figure1}(b) only the case of a measurement performed with s-polarized light is presented for simplicity. Superscripts \textit{r} and \textit{i} denote the reflected and incident light, respectively. The reflected light contains a p-polarized component which causes a change in the polarization state. The polarization state of the reflected light is then given by:
\begin{align}
\label{complex_Kerr_rotation}
\chi_s = \frac{E_p^r}{E_s^r}=\frac{r_{ps}}{r_{ss}} \approx \theta_s+i\epsilon_s \thickspace \text{or} \\ \chi_p =\frac{E_s^r}{E_p^r}=\frac{r_{sp}}{r_{pp}} \approx \theta_p+i\epsilon_p
\end{align}
where $\chi_{s(p)}$ is the complex Kerr angle, $\theta_{s(p)}$ is the Kerr rotation, and $\epsilon_{s(p)}$ is the Kerr ellipticity for \textit{s}- (\textit{p}-) incident polarized light. Optical reflectance spectroscopy have been performed for both samples, as well. The reflectance spectra were recorded by using the exact same geometry which was used for the study of the magneto-optical response, with angles $\theta_{i} = \theta_{r} = 55$~degs. Reflectance spectra for both s- and p-incident polarized light have been recorded. For a more detailed description of the experimental techniques the reader is referred to Supplemental Material.

In Fig.~\ref{Figure1} (c) (top graphs), we present the measured reflectance spectra for both incident s- ($R_{ss}$ plot) and p-polarized ($R_{pp}$ plot) light. It is apparent, that the reflectance of the sample containing no AuNPs, retains a constant value in the spectral region of 550 - 800 nm. In the case of the sample containing AuNPs, a broad drop in the $R_{ss}$ and $R_{pp}$ spectra is located in the spectral region 550 - 900 nm, reaching a minimum at $\sim$ 675 nm. The reflectance reduction in this spectral region, where the electronic transitions of Bi:YIG \cite{Doormann1984,PhysRevB.12.2777} have no relevant influence, is solely attributed to Localized Surface Plasmon (LSP) resonances in the AuNPs, for sizes smaller than $\sim$ 100 nm.

Our interest is particularly focused on the influence of the LSPs on the magneto-optical response in L-MOKE geometry. The L-MOKE geometry is easy to be implemented and therefore quite attractive for technological applications, as well. In order to achieve that, we extracted the modulus of the complex Kerr angle $\chi$, which is also called magneto-optical activity (M.O.A.), for applied external magnetic fields sufficiently large to saturate the sample along the film in-plane direction. The spectral dependence of M.O.A for s- and p- incident polarized light is shown in the bottom graphs of Fig.~\ref{Figure1}(c), for both samples. The increased M.O.A. of Bi:YIG in the low-wavelength spectral region, is attributed to the wings of the magneto-optical transitions located at photon energies of 2.8 and 3.3 eV (442 and 378 nm respectively) \cite{PhysRevB.12.2777}. It becomes apparent that the magneto-optical response of the Bi:YIG layer, for the sample containing AuNPs, is strongly modified in the very broad optical region where the resonant localized surface plasmon phenomena occur. Furthermore, by observation it can be deduced that the enhancement of the M.O.A is larger in the case of incident p-polarized light. The experimental results clearly show a significant broadband enhancement of the magneto-optical response of Bi:YIG in a spectral regime far from the inherent Bi:YIG magneto-optical transitions, which is originating from LSPs in the hybrid sample.

\begin{figure*}
  \includegraphics[width=0.8\textwidth]{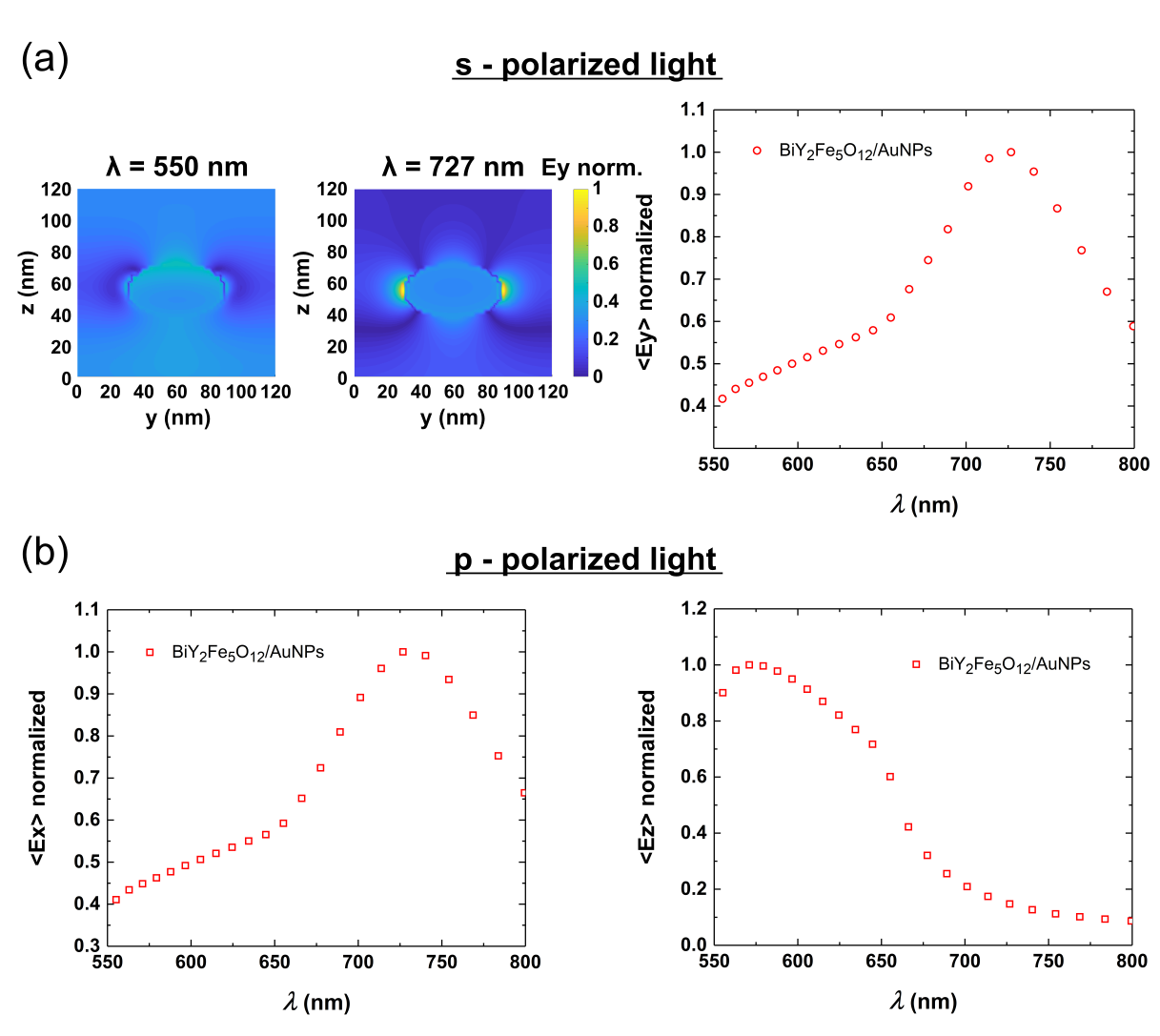}
  \caption{(a) (Left side) Maps of the y electric near-field component ($E_y$) for two different wavelength values of incident s-polarized light. The color scale has been normalized to the maximum spatial field value in each case. (Right side) Calculated $<E_y>$ in the AuNP, for incident s-polarized light. The values have been calculated from the simulated y electric near-field components in the Au nanoparticle. The $<E_y>$ values have been normalized to the maximum value. (b) Calculated $<E_x>$ and $<E_z>$ in the AuNP, for p-polarized incident light. The values have been calculated by following the same procedure as in (a).}
\label{Near_Field_simulations}
\end{figure*}

In order to gain insight into the deeper mechanism of the enhanced magneto-optical activity of the hybrid BiY$_2$Fe$_{5}$O$_{12}$/AuNPs structure, we performed numerical simulations and analyzed both the electric near-field features, as well as the far-field magneto-optical behaviour. To perform these simulation, the dielectric tensor of BiY$_2$Fe$_{5}$O$_{12}$ was used. The values of the diagonal and off-diagonal elements as a function of wavelength were taken from the literature \cite{PhysRevB.12.2777,Doormann}. For the L-MOKE geometry and the vector directions as they are defined in Fig.~\ref{Figure1} (b), the dielectric tensor of Bi:YIG has the following form:
\begin{equation}
\label{dielectric_tensor}
\varepsilon^{Bi:YIG}(\lambda) = 
\left(
\begin{array}{ccc}
  \varepsilon & 0 & 0\\
  0 & \varepsilon & \varepsilon_{yz}\\  
  0 & -\varepsilon_{yz} & \varepsilon\\
\end{array}
\right)
\end{equation}
The magneto-optical activity of the material is attributed to the off-diagonal elements of the dielectric tensor. The dielectric tensor values for Au and GGG, were also taken from literature \cite{Pt_referernce,Doormann,Palik,GGG_reference}. The angle of incidence for the incoming planar electromagnetic wave was defined at 55 $\deg$, in order to match with the angle of incidence which was used for the measurements (presented in Fig. 1(c)). The Au nanoparticles are modeled as oblate spheroids with fixed values of semi-axes: a = 60 nm and c = 35 nm. These values are chosen to correspond to the mean value of the nanoparticle sizes obtained from cross-sectional transmission electron microscopy (TEM) images \cite{Uchida2015}.

\begin{figure}
\includegraphics[width=0.5\textwidth]{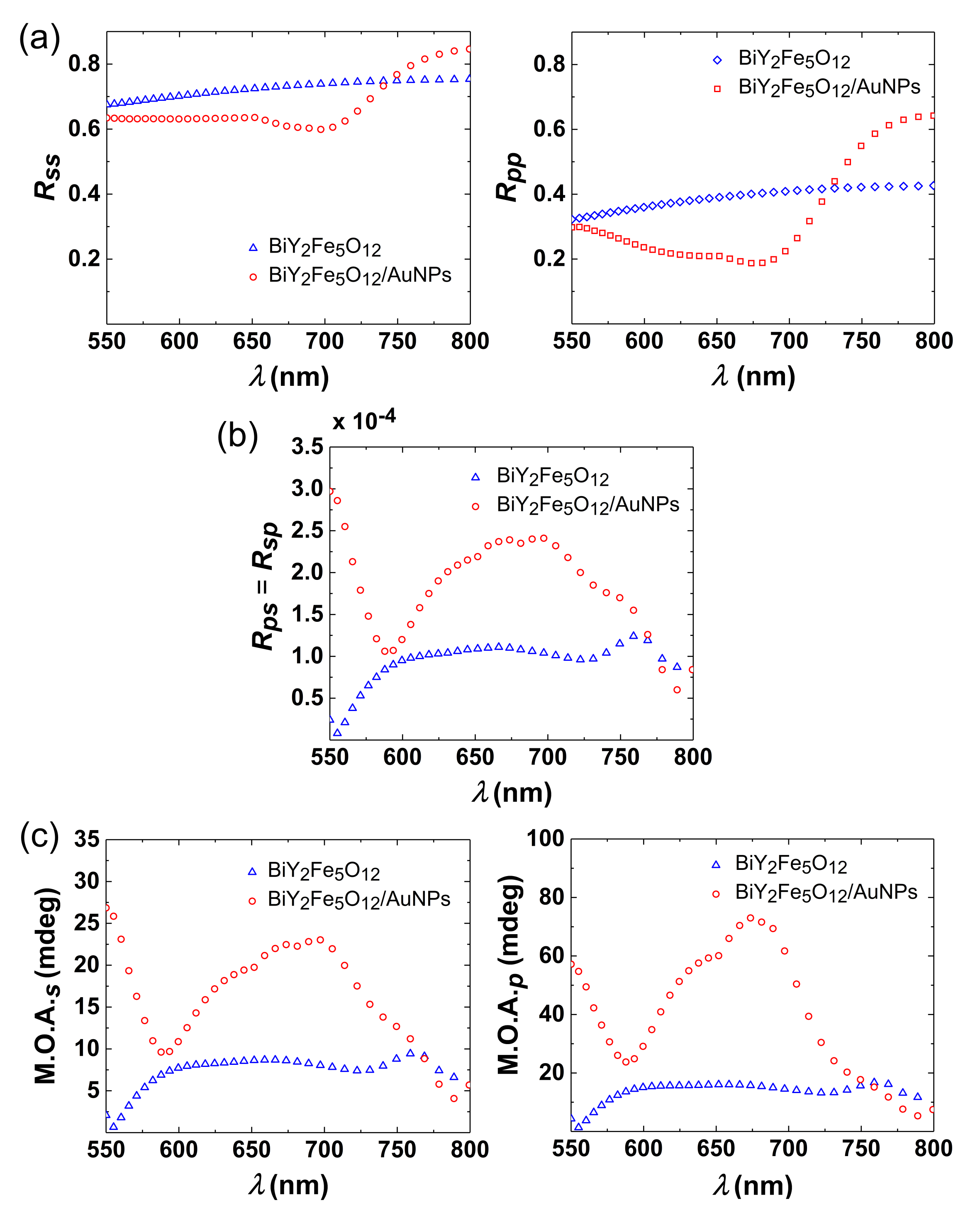}
\caption{(a) Simulated $R_{ss}$ and $R_{pp}$ for both samples. (b) Polarization conversion efficiency $R_{ps}$ or $R_{sp}$. (c) Post - calculated M.O.A.s, extracted from the simulated $r_{ss}$, $r_{pp}$, $r_{ps}$, and $r_{sp}$ values.}
\label{Floquet_simulations}
\end{figure}

In Fig.~\ref{Near_Field_simulations} (a) (left side) the maps of the spatial y component ($E_y$) of the local electric field E is presented at 2 characteristic spectral positions: $\lambda = $ 550 nm, and 727 nm. In this case the incident electric field is s-polarized. The maps reveal the spectral position of the electric near-field intensification which emanate from the localized plasmon resonances in AuNPs. As it can be observed, at $\lambda = $ 727 nm a big volume of the hosting Bi:YIG is exposed to the enhanced near-field around the AuNP. In order to quantify these near-field results in a more comprehensive way, and reveal the exact spectral positions of the plasmonic resonances, we calculate the induced dipole moments in the AuNP. The dipole moment induced in the AuNP, which is surrounded by the dielectric Bi:YIG, is given by the following formula \cite{Antonio_dielectric_antennas,Albaladejo_10}:
\begin{equation}
\label{dipole_moment}
\overrightarrow{p} = \epsilon_o(\boldsymbol{\epsilon}^{Au} - \boldsymbol{\epsilon}^{Bi:YIG})\frac{1}{N_p} \sum_{m} V_m \overrightarrow{E}_m
\end{equation}
where $\epsilon_o$ is the vacuum permittivity, $\epsilon_{Au}$ is the dielectric tensor of Au, and $\epsilon_{Bi:YIG}$ is the dielectric tensor of the surrounding Bi:YIG material. $N_p$ is the total number of the discrete mesh cells, at which each value of the electric field $\overrightarrow{E}_m$ is calculated numerically. $V_m$ is the volume of the $m^{th}$ cell of the total discretized volume. From eq. \ref{dipole_moment}, it can be deduced that $\overrightarrow{p} \propto <\overrightarrow{E}>$, where $<\overrightarrow{E}> = \frac{1}{N_p} \sum_{m} V_m \overrightarrow{E_m}$ is the mean electric field in the AuNP. The calculated y component of the mean field ($<E_y>$) as a function of the wavelength of the incident light, in the case of s-incident light, is shown in the right hand side plot of Fig. ~\ref{Near_Field_simulations} (a). The results show a clear intensification of the mean field along the y direction, with the peak of $<E_y>$ located at 727 nm. All of the values have been normalized to the maximum value of the curve. In the case of p-polarized light, the vector of the oscillating incident electric field can be analyzed along the x and the vertical z direction. Therefore, in Fig.~\ref{Floquet_simulations} (b), we choose to show both $<E_x>$ and $<E_z>$. The plots reveal the existence of a clear intensification along the x direction (where $<E_x>$ is considered), as well as a clear but more feature-complicated intensification spectrum along the z direction (where $<E_z>$ is considered). By comparison, we can deduce that the $<E_y>$ plot in the case of s-polarized incident light, and the $<E_x>$ plot in the case of p-polarized incident light, have identical shapes. This, can be clearly explained from the symmetry of the geometrical shape of the simulated nanoparticle on the xy plane. From the $<E_z>$ plot in Fig.~\ref{Floquet_simulations}(b), we can observe a field intensification close to the lower limit of the simulated spectral region. A clear peak is shown at $\lambda =$ 575 nm, with an extra feature at about $\lambda =$ 625 nm. In the case of p-polarized incident light, $<E_z>$ gets enhanced at a much smaller wavelength value than the component $<E_x>$ does. This, can be understood by the fact that the dimension of the nanoparticle along the z direction is smaller than that along the x or y direction. Therefore, the resonant mode along z is shifted at lower wavelength values, as it is compared to the resonances along the x or y direction. Furthermore, the fact that the $<E_z>$ curve has a more complex shape, in comparison to the $<E_x>$ curve, could be explained by the geometric complexity of the oblate spheroid shape.

Subsequently, we want to compare the near-field simulations with the obtained far-field data. Therefore, we initially calculated the reflectivity for s-polarized light $R_{ss}$, as well as the polarization conversion efficiency $R_{ps}$, in the L-MOKE geometry. These values are defined as follows: $R_{ss} = r_{ss}r_{ss}^\star$ and $R_{ps} = r_{ps}r_{ps}^\star$. The elements $r_{ss}$ and $r_{ps}$ have been calculated with the aid of CST by calculating the scattering matrix elements in the defined waveguide port above the simulated structure. The simulating method is based on the calculation of the power of the electromagnetic waves impinging on the defined waveguide port (see Supplemental Material for further details), and provides very useful qualitative information about the reflectivity and the polarization conversion efficiency.

The simulated $R_{ss}$ and $R_{pp}$ for the corresponding oblique angle of incidence of $\theta_i = 55 \deg$, are presented in Fig.~\ref{Floquet_simulations}(a). For incident s-polarized light ($R_{ss}$), in the case of the film containing AuNPs, the characteristic minimum in the reflectivity associated with plasmon excitation is observed at about 700 nm. Plasmon excitation is further verified by the spectral position of the maximum of the near-field $<E_y>$, (Fig. 2 (a)). Furthermore, the simulated $R_{ss}$ plot reproduces the corresponding experimental one very well. In the case of incident p-polarized light ($R_{pp}$), the modification due to the LSPs is stronger than in the case of incident s-polarized, as observed in the experimental $R_{pp}$ curve in Fig. 1 (c).

It is worthwhile to notice that the influence of the two resonances associated to $<E_x>$ and $<E_z>$ appearing at two distinct spectral positions (see in Fig.2 (b)), becomes visible in the $R_{pp}$ curve. This distinction is not clear in the experimental curves, we surmise, due to the dispersion of the aspect ratios of the Au nanoparticles and other typical imperfections in the actual samples. To account for this distribution of the sizes and the aspect ratios, a large number of simulations over different configurations would be required, leading to an unaffordable time-scale for each numerical spectrum. In Fig. 3 (b), we show the simulated polarization conversion efficiencies $R_{ps}$ or $R_{sp}$ for each sample. These appear always to be identical in every case, namely $R_{ps} = R_{sp}$. The latter is dictated by the symmetry imposed by the material itself ($\epsilon_{yz}=-\epsilon_{zy}$). The simulations show that $R_{ps(sp)}$ clearly exhibits a broadband enhancement in the spectral region of 600 - 800 nm where the plasmonic resonances are located, and it is independent of the incident light polarization. This proves that even in the case of incident s-polarized light, there is a resonance along the z direction. This resonance is the result of polarization coupling, and it is mediated through the magneto-optical properties of Bi:YIG, giving rise to the broadband enhancement of the polarization conversion. In Fig. 3 (c), the post-calculated M.O.A for the case of s- and p-polarized incident light are presented. The simulated M.O.A. reproduces very well the experimental one in the spectral region of about 675 nm, in the case of the structure containing AuNPs. By comparing the simulated M.O.A. curves, we see that the one corresponding to the pure Bi:YIG structure without AuNPs, becomes largely modified by the presence of AuNPs. In the case of incident p-polarized light, the influence of the two characteristic features attributed to the LSP resonances along the two semi-axes of the nanoparticle, are visible. The distinction between these two resonances in the experimental data (Fig.1 (c)) are, again, washed out due to the distribution of the nanoparticle sizes and their aspect ratios. The modified M.O.A. in the far-field is generated by the landscape of the electric near-field modifications. These near-field features in the plasmonic structure have a direct impact on the exhibited reflectances as well as on the polarization conversion efficiency, and therefore on the magneto-optical activity.

In summary, we have experimentally demonstrated a broad-band enhancement of the magneto-optical activity of hybrid Bi:YIG/AuNP systems induced by localized plasmon resonances, by analyzing the longitudinal magneto-optical Kerr configuration. In order to unravel the role played by the localized plasmon resonances on the magneto-optical behaviour of the host Bi:YIG, we performed near-field simulations, and correlated the obtained results with the numerical far-field spectra. We have unambiguously shown that the features in the enhanced magneto-optical activity are the result of two orthogonal LSP resonances that are coupled by the MO activity of the underlying Bi:YIG matrix, and are presented for both polarization states of the incident light. Our results pave the way to the design on-demand of the magneto-optical response of hybrid magneto-plasmonic circuitry, by controlling the localized resonances through the size and the aspect ratio of the nanoparticles.

E.Th.P. and S. D. P. acknowledge the Carl Zeiss Foundation for financial support. We gratefully acknowledge the Deutsche Forschungsgemeinschaft program SFB/TRR 173: SPIN+X Project B07. Ken-ichi Uchida (NIMS, Japan) is acknowledged for providing the Bi:YIG/AuNPs samples. Dipl.-Phys. Marc Vogel - AG von Freymann - TU Kaiserslautern is acknowledged for the introduction to CST software and his help with the installation of the package.

\bibliography{References}

\end{document}